\documentclass[12pt,twoside]{article}
\begin{document}
\baselineskip=0.20in
\vspace{20mm}
\baselineskip=0.30in
\begin{center}

{\large \bf Bound state solutions of the Dirac-Rosen-Morse potential with spin and pseudospin symmetry.}
\vspace{10mm}

{ \large \bf K.\,J. Oyewumi\footnote{E-Mail:~mjpysics@yahoo.com or kjoyewumi66@unilorin.edu.ng\\International Chair of Mathematical Physics and Applications (ICMPA-UNESCO Chair) Universite d'Abomey-Calavi, Cotonou, Republic of Benin.} and C.\,O. Akoshile\footnote{ clemakos@unilorin.edu.ng
    
} }
\vspace{5mm}

 { Theoretical Physics Section, Department of Physics\\ University of Ilorin,  P. M. B. 1515, Ilorin, Nigeria. \\

}

\baselineskip=0.20in

\vspace{4mm}

\end{center}

%\vspace{4mm}

%\maketitle
\begin{abstract}
\noindent
The energy spectra and the corresponding two- component spinor
wavefunctions of the Dirac  equation for the Rosen-Morse potential with
spin and pseudospin symmetry are obtained. The $s-$wave
($\kappa = 0$ state) solutions for this  problem are obtained by using the basic concept of the
supersymmetric quantum mechanics approach and function analysis
(standard approach) in the calculations. Under the spin symmetry
and pseudospin symmetry, the energy equation and the corresponding two-component spinor wavefunctions for this potential and other special types of this potential are obtained. Extension of
this result to $\kappa \neq 0$ state is suggested.
\end{abstract}

\baselineskip=0.28in \vspace{2mm}

\noindent
{ {\bf KEY WORDS}}: Dirac equation; Rosen-Morse potential; Pseudospin symmetry; Spin symmetry; Supersymmetric; Energy equations; Spinors; PT-symmetric Rosen-Morse potential;
Eckart-type potential; Reflectionless-type potential.

\baselineskip=0.28in \vspace{2mm}

\noindent
{\bf PACS}:  03.65.Ge, 03.65.Pm, 02.30.Gp, 11.30.Pb \vspace{2mm}

\baselineskip=0.28in \vspace{2mm}

\noindent
\section{\bf Introduction}
The Rosen-Morse potential \cite{RoM32} plays a fundamental role in
Atomic, Chemical and Molecular Physics, since it can be used to
describe molecular vibrations and to obtain the energy spectra of
linear and nonlinear systems. This potential is very useful for
describing interatomic interaction of the linear molecules and
helpful for describing polyatomic vibration energies including the
vibration states of $NH_{3}$ molecule.

The Rosen-Morse potential \cite{RoM32} is given as
\begin{equation}
V(r)= -V_{1}  \textrm{sech}^{2}{\alpha r} + V_{2}\tanh{\alpha r} 
\label{r1}
\end{equation}
where $V_{1}$ and $V_{2}$ are the depth of the potential and
$\alpha$ is the range of the potential, respectively. The bound
state energy eigenvalues and the corresponding wavefunctions of the
``deformed'' Rosen-Morse and modified Rosen-Morse potentials have
been obtained by using Nikiforov-Uvarov method \cite{BuE97, EgE99}.
For this same potential, path integrals approach has been used to solve the Schr\"{o}dinger
equation \cite{GrS98}. Also, it has been shown that the
Rosen-Morse potential and its PT-symmetric version are the special
cases of the five-parameter exponential-type potential model
\cite{JiE02,JiE03}. 

In addition, the energy equations and wavefunctions
of the relativistic Klein-Gordon equation for the Rosen-Morse well
and their PT-symmetric versions which are the special cases of the
Rosen-Morse-type potential have been obtained \cite{YiE04}.
The exact solutions of the trigonometric Rosen-Morse potential have been obtained by using the supersymmetric
\cite{CoK06} and the improved quantization rule methods \cite{MaE07}. Approximate solutions of the
Schr\"{o}dinger equation for the Rosen-Morse potential for all
values of orbital angular momentum quantum number (including
centrifugal term $\ell \neq 0$) have been obtained \cite{Tas09}.

The solution of the Dirac equation with mixed potentials
for particles such as atoms, nuclei and hadrons play a central role
in a realistic nuclear system \cite{Lev04, LeV05}. The essence of the relativistic descriptions introduced, is to understand the relativistic behaviour of spin $1/2$ particles. In order to
investigate the nuclear shell structure, the study of pseudospin and
spin symmetric solutions of the Dirac equation has been an important
area of research in nuclear physics
[13 - 30].
%\cite{ArE69,HeA69,BoE82,DuE87,MeR96,Gin97,GiM98,AlE01, AlE02,LiE041,LiE042,LiE043,GuE051,GuE052,GuF06,Gin04, Gin051,Gin052}.

The idea about pseudospin and spin symmetry with the nuclear shell
model has been introduced $5$ decades ago \cite{ArE69, HeA69}, this
idea  has been widely used in explaining a number of phenomena in
nuclear physics and related areas.
The spin symmetry in Dirac equation occurs when the difference of
the potential between the vector potential $V(r)$ and scalar potential
$S(r)$ is a constant ({\it i. e.} , $\Delta(r) = V(r)- S(r) $= constant),
and the pseudospin symmetry occurs when the sum of the potential of the
vector potential $V(r)$ and scalar potential $S(r)$ is a constant
({\it i. e.} , $\Sigma(r) = V(r) + S(r)$ = constant)
\cite{Gin97,Gin04, Gin051,Gin052}. 

These have been investigated in several nuclei
for a few potentials in many areas of physics.
For instance, the spin symmetry is relevant for mesons \cite{PaE02};
the pseudospin symmetry in nuclear theory refers to quasi-degeneracy
of single-nucleon doublets and can be characterized in the
non-relativistic quantum numbers ($n, \ell, j= \ell + \frac{1}{2}$) and
($n - 1, \ell + 2, j = \ell + \frac{3}{2} $), where $n$, $\ell$ and
$j$ are the single-nucleon radial, orbital, and total angular
momentum quantum numbers, respectively \cite{ArE69, HeA69}.  These
levels have the same pseudo-orbital angular momentum quantum number,
$ \overline{\ell} = \ell + 1  $, pseudospin quantum number, $
\overline{s} = \frac{1}{2} $. Pseudospin symmetry is exact when
doublets with $j = \overline{\ell} \pm \overline{s}$ are degenerate
\cite{LiE041,LiE042,LiE043,SuE99}.

Some potentials of interest are solved ranging from exact analytical solutions
(with various methods) to approximate analytical methods. These
potentials include: harmonic oscillator [22-26, 33] %\cite{LiE041,LiE042,LiE043,GuE051,GuE052,CaE04};
Coulomb potential \cite{AlE06, Thy09, Akr09};
Hulth$\acute{e}$n potential \cite{SoE07,SoE081,SoE082}; Morse
potential [40 - 45]. 
%\cite{Ber06, Ber09, QiE07,BaB07,ZhE08,ZhE09}.
Others are: Woods-Saxon potential [22 - 24, 46, 47] %\cite{AlE01,AlE02,LiE041,LiE042,LiE043,GuS05,XuZ06}
; Eckart potential
\cite{SoE07,SoE081,SoE082,JiE06};
Pseudoharmonic potential \cite{AyS09};
generalized asymmetrical Hartmann potential \cite{GuE07};
Manning-Rosen potential \cite{Tas09,WeD08,ChE09}.

In addition, the Dirac equation with spin and pseudospin symmetry
for the following potentials have been solved also: generalized
P\"{o}schl-Teller and hyperbolic potentials \cite{XuE08,JiE091,JiE092};
 ring-shaped non-spherical harmonic oscillator \cite{Zha091}; double ring-shaped spherical harmonic oscillator potential
\cite{Zha092}; Kratzer potential with an angle-dependent potential \cite{BeS08}.

Recently, by using Nikiforov-Uvarov method, the Dirac equation is solved for some exponential potentials (the hypergeometric-type potential, the generalized Morse potential, and the P\"{o}schl-Teller potential) with any spin-orbit
quantum number $\kappa$ in the case of spin and pseudospin symmetry \cite{ArE10}. Using Nikiforov-Uvarov method also, exact solutions of the Dirac equation with the Mie-type potential under pseudospin and spin symmetry limit have been obtained \cite{AyS10}.

Motivated by the success in obtaining the bound state solutions for: 
the Schr\"{o}dinger equations for the Rosen-Morse potential with and
without centrifugal term ({\it i. e.} , $\ell=0$ and $\ell \neq 0$) [1 - 10] and the Dirac equations for
the Rosen-Morse potential with and without centrifugal term ({\it i. e.} , $\ell=0$ and $\ell \neq 0$) \cite{JiE02, JiE03,YiE04}. We attempt to study the bound state solutions for the
Rosen-Morse potential with spin and pseudospin symmetry in terms of
the supersymmetric shape invariance formalism and function analysis
method.

This paper is organized as follows: In Section $2$, we present the basic equations for the associated two-component spinors of the Dirac equation. Section $3$ is devoted to finding the bound state solutions of the Dirac-Rosen-Morse potential, the spin symmetry and pseudospin symmetry solutions are obtained. In Section $4$, we study some special cases of the Rosen-Morse type potential under the spin symmetry and pseudospin symmetry conditions. We summarize our conclusion in Section $5$.

\section{Basic equation for the associated two-component spinors of the Dirac equation.}
Using Dirac wave equation and Dirac spinor wave functions, we can obtain the two-coupled second-order ordinary differential equations for the upper and lower components of the Dirac wavefunction as follows:
\begin{equation}
\left(\frac{d}{dr} + \frac{\kappa}{r} \right)F_{n \kappa}(r) = \left[ M + E_{n \kappa} + S(r) - V(r)\right]G_{n \kappa}
\label{r2},
\end{equation}
\begin{equation}
\left(\frac{d}{dr} - \frac{\kappa}{r} \right)G_{n \kappa}(r) = \left[ M - E_{n \kappa} + S(r) + V(r)\right]F_{n \kappa}
\label{r3}.
\end{equation}

Eliminating $G_{n \kappa}(r)$ in eq.(2) and $F_{n \kappa}(r)$ in eq.(3), we obtain, the following two second-order differential equations for the upper and lower components:
\begin{equation}
\left(- \frac{d^{2}}{dr^{2}} + \frac{\ell(\ell + 1)}{r^{2}} + (M + E_{n \kappa} - \Delta(r))(M - E_{n \kappa} + \Sigma(r)) -\frac{\frac{d \Delta(r)}{dr} (\frac{d}{dr} + \frac{\kappa}{r} )}{M + E_{n \kappa} - \Delta(r)} \right)F_{n \kappa}(r)=0
\label{r4},
\end{equation}
\begin{equation}
\left(- \frac{d^{2}}{dr^{2}} + \frac{ \overline{\ell}(\overline{\ell} + 1)}{r^{2}} + (M + E_{n \kappa} - \Delta(r))(M - E_{n \kappa} + \Sigma(r)) +\frac{\frac{d \Sigma(r)}{dr} (\frac{d}{dr} - \frac{\kappa}{r} )}{M - E_{n \kappa} + \Sigma(r)}    \right)G_{n \kappa}(r)=0
\label{r5},
\end{equation}
where $\Delta(r) = V(r) - S(r)$ and $\Sigma(r) = V(r) + S(r)$ are the difference and the sum of the potentials $V(r)$ and $S(r)$, respectively. We consider bound state solutions that demand the radial components satisfying $F_{n \kappa}(0)=G_{n \kappa}(0)=0$ and $F_{n \kappa}(\infty)=G_{n \kappa}(\infty)=0$.

\section{Bound state solutions of the Dirac-Rosen-Morse potential.}

\subsection{Spin symmetry solutions of the Dirac-Rosen-Morse potential.}

Under the condition of the spin symmetry, {\it i. e.} , $\frac{d \Delta(r)}{dr} = 0$ or $\Delta(r) = C$ (constant), eq. (\ref{r4}) reduces into
\begin{equation}
\left[- \frac{d^{2}}{dr^{2}} + \frac{\ell(\ell + 1)}{r^{2}} + (M + E_{n \kappa} - C_{s}) \Sigma(r)  \right]F_{n \kappa}(r) =\left[E_{n \kappa}^{2} - M^{2} + C_{s} (M - E_{n \kappa}) \right]F_{n \kappa}(r)
\label{r6}.
\end{equation}
We take the Rosen-Morse potential as the $\Sigma(r)$ [1 - 10]
%\cite{RoM32,BuE97, EgE99, GrS98, JiE02, JiE03, YiE04, CoK06, MaE07, Tas09},
\begin{equation}
\Sigma(r) = -V_{1} \textrm{sech}^{2} {\alpha r} + V_{2}~ \tanh {\alpha r}
\label{r7}.
\end{equation}
Substituting eq.(\ref{r7}) into eq.(\ref{r6}), we obtain a Schr\"{o}dinger-like equation for the $s$-wave ($\ell = 0$, {\it i. e.} $\kappa = -1$),
\begin{equation}
\left[- \frac{d^{2}}{dr^{2}} + \widetilde{V_{1}} \textrm{sech}^{2} {\alpha r} + \widetilde{V_{2}}~ \tanh {\alpha r } \right]F_{n, -1}(r) =\widetilde{E} F_{n, -1}(r)
\label{r8},
\end{equation}
where $ \widetilde{V_{1}} = -V_{1} (M + E_{n, -1} - C_{s}), ~\widetilde{V_{2}} = V_{2} (M + E_{n, -1} - C_{s}),~ \widetilde{E} = E_{n, -1}^{2} - M^{2} + C_{s} (M - E_{n, -1}).$

We employ the basic concept of the supersymmetric quantum mechanics method and shape invariance approach to solve eq. (\ref{r8}) [61 - 65]. 
%\cite{Wit81, Gen83, CoE95, Suk05,CoE05}.
The ground-state function for the upper radial component $F_{n \kappa}$ can be written in the form
\begin{equation}
F_{0, -1}(r) exp(-\int W(r)dr)
\label{r9},
\end{equation}
where $W(r)$ is called a superpotential in supersymmetry quantum mechanics [61 - 65]. 
%\cite{Wit81, Gen83, CoE95, Suk05,CoE05}.
On substituting eq. (\ref{r9}) into eq. (\ref{r8}), we obatin the following equation for $W(r)$,
\begin{equation}
W^{2}(r) -\frac{dW(r)}{dr} = \widetilde{V_{1}} \textrm{sech}^{2} {\alpha r} + \widetilde{V_{2}}~ \tanh {\alpha r} - \widetilde{E}_{0} = \frac{4 \widetilde{V}_{1}e^{2\alpha r} }{(e^{2\alpha r + 1})^{2}} + \frac{\widetilde{V}_{2}(e^{2\alpha r} - 1) }{(e^{2\alpha r + 1})} - \widetilde{E}_{0}
\label{r10},
\end{equation}
eq. (\ref{r10}) is a non-linear Riccati's equation.

In order to make the superpotential $W(r)$ be compatible with the property of the right hand side of eq. (\ref{r10}),  we write the superpotential $W(r)$ as
\begin{equation}
W^{2}(r) = Q_{1} + \frac{Q_{2}}{e^{2\alpha r} + 1} e^{2\alpha r}
\label{r11}.
\end{equation}
We obtain the ground-state function $F_{0, -1}(r)$
\begin{equation}
\displaystyle{F_{0, -1}(r) = e^{-Q_{1} r}(e^{2 \alpha r} + 1)^{-\frac{Q_{2}}{2 \alpha}}}
\label{r12},
\end{equation}
for the bound state solutions that demand the upper component $F_{n \kappa}$ satisfying the regularity conditions $F_{n \kappa}(0) = F_{n \kappa}(\infty) = 0$. These boundary conditions allow us to obtain the restriction conditions $Q_{1}>0$ and $Q_{2}<0$. By these restriction conditions, we obtain
\begin{equation}
Q_{1} = \frac{\widetilde{V}_{2}}{Q_{2}} - \frac{Q_{2}}{2}
\label{r13}
\end{equation}
and
\begin{equation}
Q_{2} = -2 \alpha \beta
\label{r14},
\end{equation}
where parameter $\beta$ is defined as
\begin{equation}
\beta = \frac{1}{2}\left(-1 + \sqrt{1 + \frac{4 V_{1}(M + E_{n, -1} - C_{s}) }{\alpha^{2}}}\right)
\label{r15}.
\end{equation}
The ground-state energy and superpotential can be expressed as
\begin{equation}
\widetilde{E}_{0} = - \left[\frac{\widetilde{V}_{2}}{Q_{2}} - \frac{Q_{2}}{2}\right ]^{2} -\widetilde{V}_{2}
\label{r16}
\end{equation}
and
\begin{equation}
W(r) =  \left[\frac{\widetilde{V}_{2}}{Q_{2}} - \frac{Q_{2}}{2}\right ]  +\frac{Q_{2}}{e^{2 \alpha r} + 1}e^{2 \alpha r}
\label{r17}.
\end{equation}

By using the superpotential $W(r)$ given in eq. (\ref{r17}), we can construct the following two supersymmetry partner potentials as follows:
\begin{equation}
U_{+}(r) = W^{2}(r) + \frac{dW(r)}{dr}=  \left[\frac{\widetilde{V}_{2}}{Q_{2}} - \frac{Q_{2}}{2}\right ]^{2}  + \frac{(2Q_{1}Q_{2} - Q_{2}^{2})}{e^{2 \alpha r} + 1}e^{2 \alpha r} - \frac{(Q_{2}^{2} + 2 \alpha Q_{2})}{(e^{2 \alpha r} + 1)^{2}}e^{2 \alpha r}
\label{r18},
\end{equation}
\begin{equation}
U_{-}(r) = W^{2}(r) - \frac{dW(r)}{dr}=  \left[\frac{\widetilde{V}_{2}}{Q_{2}} - \frac{Q_{2}}{2}\right ]^{2}  + \frac{(2Q_{1}Q_{2} + Q_{2}^{2})}{e^{2 \alpha r} + 1}e^{2 \alpha r} - \frac{(Q_{2}^{2} - 2 \alpha Q_{2})}{(e^{2 \alpha r} + 1)^{2}}e^{2 \alpha r}
\label{r19}.
\end{equation}
By these two supersymmetry partner potentials, we obtain the following shape invariance relationship satisfied by the partner potentials $U_{-}(r)$ and $U_{+}(r)$:
\begin{equation}
U_{+}(r, a_{0}) = U_{-}(r, a_{1}) + R(a_{1})
\label{r20},
\end{equation}
where $a_{0} = Q_{2}$, $a_{1}$ is a function of $a_{0}$, that is, $a_{1} = f(a_{0}) = Q_{2} -2 \alpha $, and the remainder $R(a_{1})$ is independent of $r$, $R(a_{1}) = \left[\frac{\widetilde{V}_{2}}{a_{0}} - \frac{a_{0}}{2}\right ]^{2} - \left[\frac{\widetilde{V}_{2}}{a_{1}} - \frac{a_{1}}{2}\right ]^{2} $.

Eq. (\ref{r20}) shows that the two partner potentials $U_{-}(r)$ and $U_{+}(r)$ have similar shapes and they are shape-invariant potentials \cite{Gen83}. By employing shape invariance approach [61 - 65], 
%\cite{Wit81, Gen83, CoE95, Suk05,CoE05},
we can calculate exactly the energy spectrum of the shape-invariant potential $U_{-}(r)$ and obtain the following results:
\begin{equation}
\widetilde{E}_{0}^{(-)} = 0
\label{r21},
\end{equation}
\begin{eqnarray}
\widetilde{E}_{n}^{(-)} &= \sum_{k = 1}^{n} R(a_{k})= R(a_{1}) + R(a_{2}) + \ldots +R (a_{n})\\ \nonumber
& = \left[\frac{\widetilde{V}_{2}}{a_{0}} - \frac{a_{0}}{2}\right ]^{2} - \left[\frac{\widetilde{V}_{2}}{a_{1}} - \frac{a_{1}}{2}\right ]^{2} + \left[\frac{\widetilde{V}_{2}}{a_{1}} - \frac{a_{1}}{2}\right ]^{2} - \left[\frac{\widetilde{V}_{2}}{a_{2}} - \frac{a_{2}}{2}\right ]^{2} + \ldots \\ \nonumber
& \left[\frac{\widetilde{V}_{2}}{a_{n -1}} - \frac{a_{n -1}}{2}\right ]^{2} - \left[\frac{\widetilde{V}_{2}}{a_{n}} - \frac{a_{n}}{2}\right ]^{2}\\ \nonumber
& =\left[\frac{\widetilde{V}_{2}}{a_{0}} - \frac{a_{0}}{2}\right ]^{2} - \left[\frac{\widetilde{V}_{2}}{a_{n}} - \frac{a_{n}}{2}\right ]^{2}\\ \nonumber
& \left[\frac{\widetilde{V}_{2}}{Q_{2}} - \frac{Q_{2}}{2}\right ]^{2} - \left[\frac{\widetilde{V}_{2}}{Q_{2} - 2n \alpha} - \frac{Q_{2} - 2n \alpha}{2}\right ]^{2}
\label{r22},
\end{eqnarray}
where the quantum number $n = 0, 1, 2, \ldots$.

We find that, the solution for $\widetilde{E}$ in  eq. (\ref{r8}) can be obtained as
\begin{equation}
\widetilde{E} = \widetilde{E}_{n}^{(-)} +  \widetilde{E}_{0} = - \left[\frac{\widetilde{V}_{2}}{Q_{2} - 2n \alpha}\right ]^{2} - \frac{(Q_{2} - 2n \alpha)^{2}}{4}
\label{r23}.
\end{equation}
Since $\widetilde{E} = E_{n, -1}^{2} - M^{2} + C_{s} (M - E_{n, -1})$, then , substituting eq. (\ref{r14}) into (\ref{r23}), we obtain the energy equation for the Rosen-Morse potential with spin symmetry in the Dirac theory
\begin{equation}
M^{2} - E_{n, -1}^{2}  - C_{s} (M - E_{n, -1}) = \frac{(M + E_{n, -1}  - C_{s})^{2}V_{2}^{2}}{4 \alpha^{2} (\beta - n)^{2}} +\alpha^{2} (\beta - n)^{2}
\label{r24},
\end{equation}
where the parameter $\beta$ is as given in equation (\ref{r15}).

In order to obtain the unnormalized excited wavefunctions, we use the standard function analysis methods to solve eq. (\ref{r8}). Using the energy spectrum expression in eq. (\ref{r24}), we re-write eq. (\ref{r8}) in the form:
\begin{equation}
\left[\frac{d^{2}}{dr^{2}} - \widetilde{V}_{1} \textrm{sech}^{2} {\alpha r} - \widetilde{V}_{2}~ \tanh {\alpha r } \right]F_{n, -1}(r) = \left[\frac{(M + E_{n, -1}  -C_{s})^{2}V_{2}^{2}}{4 \alpha^{2} (\beta - n)^{2}} +\alpha^{2} (\beta - n)^{2} \right] F_{n, -1}(r)
\label{r25}.
\end{equation}
By defining a new variable of the form $x= -e^{-2 \alpha r}$, we can transform eq. (\ref{r25}) into the following form:
\begin{equation}
\left[x^{2}\frac{d^{2}}{dx^{2}} + x \frac{d}{dr} + \widetilde{V}_{1} \frac{x}{\alpha^{2}(1 - x )^{2}} - \widetilde{V}_{2} \frac{(1 + x)}{\alpha^{2}(1 - x)}  - \frac{(M + E_{n, -1}  - C_{s})^{2}V_{2}^{2}}{16 \alpha^{4} (\beta - n)^{2}} -  (\beta - n)^{2} \right] F_{n, -1}(r)=0
\label{r26}.
\end{equation}

By taking the function $F_{n, -1}(x)$ as $F_{n, -1}(x) = x^{\lambda} (1 - x)^{\beta}F(x)$, we obtain eq. (\ref{r26}) in the form of the hypergeometric differential equation \cite{AbS70}
\begin{equation}
x(1 - x)\frac{d^{2}F(x)}{dx^{2}} + [2 \lambda + 1 - (2 \lambda + 2 \beta + 1)x] \frac{d F(x)}{dr} - (2 \beta + n)(2 \lambda - n) =0
\label{r27},
\end{equation}
where $\lambda = \frac{1}{2}[n + \beta + \frac{(M + E_{n, -1}  - C_{s})V_{2}}{2 \alpha^{2} (\beta - n)}]$.
Hence, the upper spinor component $F_{n, -1}(r)$ of the radial wavefunction corresponding to energy level $E_{n, -1}$ is,
\begin{equation}
F_{n, -1} (r)=  N_{n, -1}(-1)^{\lambda}e^{-2\alpha \lambda r}(1 + e^{-2\alpha \lambda r})^{\beta}~~ _{2}F_{1}(2 \beta + n, 2 \lambda - n, 2 \lambda + 1; -e^{-2\alpha \lambda r})
\label{r28}.
\end{equation}

The lower spinor component $G_{n, -1}(r)$ can be obtained from eq. (\ref{r2}) with $\kappa = -1$ and on re-writing eq. (\ref{r2}), we have,
\begin{equation}
G_{n, -1} (r)= \frac{1}{M + E_{n, -1} - C}\left[\frac{d}{dr} - \frac{1}{r} \right]F_{n, -1} (r)
\label{r29}.
\end{equation}
In order to obtain the lower spinor component $G_{n, -1}(r)$, we employ the use of the recurrence relation of the hypergeometric function \cite{AbS70},
\begin{equation}
\displaystyle{\frac{d}{d \xi} [~ _{2}F_{1}(a, b, c; \xi) ] =  \frac{ab}{c}~~_{2}F_{1}(a + 1, b + 1, c + 1; \xi)}
\label{r30}.
\end{equation}
Therefore, the lower spinor component $G_{n, -1}(r)$ corresponding to the upper component $F_{n, - 1}(r)$ and energy level $E_{n, -1}$ is,
\begin{eqnarray}
&\displaystyle{G_{n, -1} (r)= -N_{n, -1}(-1)^{\lambda}\frac{e^{-2\alpha \lambda r}(1 + e^{-2\alpha r})^{\beta}}{M + E_{n, -1} - C}}  \\ \nonumber
& \times [  \left( 2 \alpha \lambda + 2 \alpha \beta e^{-2\alpha r}(1 + e^{-2\alpha r})^{-1}  + \frac{1}{r}\right)
 ~_{2} F_{1}(2 \beta + n, 2 \lambda - n, 2 \lambda + 1; -e^{-2\alpha r})\\ \nonumber
& - \frac{2 \alpha (2 \beta + n)(2 \lambda - n) e^{-2\alpha r}}{2 \lambda + 1}~~ _{2}F_{1}(2 \beta + n + 1, 2 \lambda - n + 1, 2 \lambda + 2; -e^{-2\alpha r}) ]
\label{r31}.
\end{eqnarray}

\subsection{Pseudospin symmetry solutions of the Dirac-Rosen-Morse potential.}

For the case of exact pseudospin symmetry, {\it i. e.} , $\frac{d \Sigma(r)}{dr} = 0$ or $\Sigma(r) = C_{ps}$ (constant), eq. (\ref{r5}) reduces into
\begin{equation}
\left[- \frac{d^{2}}{dr^{2}} + \frac{\overline{\ell}(\overline{\ell} + 1)}{r^{2}} - (M - E_{n \kappa} + C_{ps}) \Delta(r)  \right]G_{n \kappa}(r) =\left[E_{n \kappa}^{2} - M^{2} - C_{ps} (M + E_{n \kappa}) \right]G_{n \kappa}(r)
\label{r32}.
\end{equation}
We take the Rosen-Morse potential as the $\Delta(r)$,
\begin{equation}
\Delta(r) = -V_{1} \textrm{sech}^{2} {\alpha r} + V_{2}~ \tanh {\alpha r}
\label{r33}.
\end{equation}
Substituting eq. (\ref{r33}) into eq. (\ref{r32}), we obtain a Schr\"{o}dinger-like equation for the $s$-wave ($\overline{\ell} = 0$, {\it i. e.} $\kappa = 1$),
\begin{equation}
\left[- \frac{d^{2}}{dr^{2}} + \widetilde{V}_{1} \textrm{sech}^{2} {\alpha r} + \widetilde{V}_{2}~ \tanh {\alpha r } \right]G_{n, 1}(r) =\widetilde{E} G_{n, 1}(r)
\label{r34},
\end{equation}
where $$ \widetilde{V}_{1} = V_{1} (M - E_{n, 1} + C_{ps}), \widetilde{V}_{2} = - V_{2} (M - E_{n, 1} + C_{ps}), \widetilde{E} = E_{n, 1}^{2} - M^{2} - C_{ps} (M + E_{n, 1}). $$
Using the same procedure of solving eq. (\ref{r8}), we obtain the energy equation for the Rosen-Morse potential with pseudospin symmetry in the Dirac theory,
\begin{equation}
M^{2} - E_{n, 1}^{2}  + C_{ps} (M + E_{n, 1}) = \frac{(M - E_{n, 1}  + C_{ps})^{2}V_{2}^{2}}{4 \alpha^{2} (\beta - n)^{2}} +\alpha^{2} (\beta - n)^{2}
\label{r35},
\end{equation}
where the parameter $\beta$ is as given in eq. (\ref{r15}) and the quantum number $n = 1, 2, 3, \ldots$. The unnormalized lower radial wavefunction is given by
\begin{equation}
G_{n, 1} (r)= N_{n, 1}(-1)^{\lambda}e^{-2 \alpha \lambda r}(1 + e^{-2\alpha  r})^{\beta}~~ _{2}F_{1}(2 \beta + n, 2 \lambda - n, 2 \lambda + 1; -e^{-2\alpha r})
\label{r36},
\end{equation}
where $\lambda$ and $\beta$ are given as, $\lambda = \frac{1}{2}[n + \beta - \frac{(M - E_{n, 1} + C_{ps})V_{2}}{2 \alpha^{2} (\beta - n)}]$ and $\beta = \frac{1}{2}\left[-1 + \sqrt{1 - \frac{4 (M - E_{n, 1} + C_{ps})V_{1} }{\alpha^{2}}}\right]$, respectively.
In a similar manner, we obtain the upper spinor component $F_{n, 1}(r)$ corresponding to the lower component $G_{n, 1}(r)$ and energy level $E_{n, 1}$ as
\begin{eqnarray}
&\displaystyle{F_{n, 1} (r)= N_{n, 1}(-1)^{\lambda}\frac{e^{-2\alpha \lambda r}(1 + e^{-2\alpha r})^{\beta}}{M - E_{n, 1} + C_{ps}}}  \\ \nonumber
& \times [  \left( -2 \alpha \lambda + 2 \alpha \beta e^{-2\alpha r}(1 + e^{-2\alpha r})^{-1}  + \frac{1}{r}\right)
 ~_{2} F_{1}(2 \beta + n, 2 \lambda - n, 2 \lambda + 1; -e^{-2\alpha r})\\ \nonumber
& - \frac{2 \alpha (2 \beta + n)(2 \lambda - n) e^{-2\alpha r}}{2 \lambda + 1}~~ _{2}F_{1}(2 \beta + n + 1, 2 \lambda - n + 1, 2 \lambda + 2; -e^{-2\alpha r}) ]
\label{r37}.
\end{eqnarray}
\section{Some special cases of the Rosen-Morse type potentials}

\subsection{Standard Eckart potential}
In this subsection, we set $V_{1} = -V_{1}$ and $V_{2} = -V_{2}$ in eq. (\ref{r1}), and eq. (\ref{r1}) reduces to the standard Eckart potential \cite{Eck30}
\begin{equation}
V(r)= V_{1}  \textrm{sech}^{2}{\alpha r} - V_{2}\tanh{\alpha r}. 
\label{r38}
\end{equation}
For the spin symmetry solutions of this potential, the energy eigenvalues, $E_{n, -1}$, the upper and lower spinor components $F_{n, -1}$ and $G_{n, -1}$ of the radial wavefunctions corresponding to this energy level are given, respectively as :   
\begin{equation}
M^{2} - E_{n, -1}^{2}  - C_{s} (M - E_{n, -1}) = \frac{(M + E_{n, -1}  - C_{s})^{2}V_{2}^{2}}{4 \alpha^{2} (\beta - n)^{2}} +\alpha^{2} (\beta - n)^{2}
\label{r39},
\end{equation}
\begin{equation}
F_{n, -1} (r)=  N_{n, -1}(-1)^{\lambda}e^{-2\alpha \lambda r}(1 + e^{-2\alpha \lambda r})^{\beta}~~ _{2}F_{1}(2 \beta + n, 2 \lambda - n, 2 \lambda + 1; -e^{-2\alpha \lambda r})
\label{r40},
\end{equation}
\begin{eqnarray}
&\displaystyle{G_{n, -1} (r)= -N_{n, -1}(-1)^{\lambda}\frac{e^{-2\alpha \lambda r}(1 + e^{-2\alpha r})^{\beta}}{M + E_{n, -1} - C}}  \\ \nonumber
& \times [  \left( 2 \alpha \lambda + 2 \alpha \beta e^{-2\alpha r}(1 + e^{-2\alpha r})^{-1}  + \frac{1}{r}\right)
 ~_{2} F_{1}(2 \beta + n, 2 \lambda - n, 2 \lambda + 1; -e^{-2\alpha r})\\ \nonumber
& - \frac{2 \alpha (2 \beta + n)(2 \lambda - n) e^{-2\alpha r}}{2 \lambda + 1}~~ _{2}F_{1}(2 \beta + n + 1, 2 \lambda - n + 1, 2 \lambda + 2; -e^{-2\alpha r}) ]
\label{r41},
\end{eqnarray}
where parameters $\lambda$ and $\beta$ become
\begin{equation}
\lambda = \frac{1}{2}[n + \beta - \frac{(M + E_{n, -1}  - C_{s})V_{2}}{2 \alpha^{2} (\beta - n)}],
\label{r42}
\end{equation}
and
\begin{equation}
\beta = \frac{1}{2}\left(-1 + \sqrt{1 - \frac{4 V_{1}(M + E_{n, -1} - C_{s}) }{\alpha^{2}}}\right)
\label{r43}.
\end{equation}

Similarly, for the pseudospin symmetry solutions of this potential, the energy eigenvalues, $E_{n, 1}$, the lower and upper spinor components $G_{n, 1}$ and $F_{n, 1}$ of the radial wavefunctions corresponding to this energy level are given, respectively as :   
\begin{equation}
M^{2} - E_{n, 1}^{2}  + C_{ps} (M + E_{n, 1}) = \frac{(M - E_{n, 1}  + C_{ps})^{2}V_{2}^{2}}{4 \alpha^{2} (\beta - n)^{2}} +\alpha^{2} (\beta - n)^{2}
\label{r44},
\end{equation}
\begin{equation}
G_{n, 1} (r)= N_{n, 1}(-1)^{\lambda}e^{-2 \alpha \lambda r}(1 + e^{-2\alpha  r})^{\beta}~~ _{2}F_{1}(2 \beta + n, 2 \lambda - n, 2 \lambda + 1; -e^{-2\alpha r})
\label{r45},
\end{equation}
\begin{eqnarray}
&\displaystyle{F_{n, 1} (r)= N_{n, 1}(-1)^{\lambda}\frac{e^{-2\alpha \lambda r}(1 + e^{-2\alpha r})^{\beta}}{M - E_{n, 1} + C_{ps}}}  \\ \nonumber
& \times [  \left( -2 \alpha \lambda + 2 \alpha \beta e^{-2\alpha r}(1 + e^{-2\alpha r})^{-1}  + \frac{1}{r}\right)
 ~_{2} F_{1}(2 \beta + n, 2 \lambda - n, 2 \lambda + 1; -e^{-2\alpha r})\\ \nonumber
& - \frac{2 \alpha (2 \beta + n)(2 \lambda - n) e^{-2\alpha r}}{2 \lambda + 1}~~ _{2}F_{1}(2 \beta + n + 1, 2 \lambda - n + 1, 2 \lambda + 2; -e^{-2\alpha r}) ]
\label{r46},
\end{eqnarray}
where $\lambda$ and $\beta$ are given as, 
\begin{equation}
\lambda = \frac{1}{2}[n + \beta + \frac{(M - E_{n, 1} + C_{ps})V_{2}}{2 \alpha^{2} (\beta - n)}]
\label{r47}
\end{equation}
and 
\begin{equation}
\beta = \frac{1}{2}\left[-1 + \sqrt{1 + \frac{4 (M - E_{n, 1} + C_{ps})V_{1} }{\alpha^{2}}}\right]
\label{r48}.
\end{equation}

\subsection{PT-symmetric Rosen-Morse potential}

If we choose $V_{2} \rightarrow - i V_{2}$ in eq. (\ref{r1}), then, eq. (\ref{r1}) becomes the PT-symmetric Rosen-Morse potential\cite{YiE04,JiE02,Tas09}
\begin{equation}
V(r)= - V_{1}  \textrm{sech}^{2}{\alpha r} + i V_{2}\tanh{\alpha r}. 
\label{r49}
\end{equation}
For a given potential $V(r)$, if $V(r) = V^{*}(-r)$ exist, the potential $V(r)$ is said to be PT-symmetric, where $P$ denotes parity operator (space reflection, $P: r\rightarrow -r$, or $ r\rightarrow \eta - r$) and $T$ denotes time-reversal operator ($ T: i\rightarrow -i$).

For the case of the spin symmetry solutions of this PT-symmetric version of the Rosen-Morse potential, the energy eigenvalues, $E_{n, -1}$ becomes
\begin{equation}
M^{2} - E_{n, -1}^{2}  - C_{s} (M - E_{n, -1}) = - \frac{(M + E_{n, -1}  - C_{s})^{2}V_{2}^{2}}{4 \alpha^{2} (\beta - n)^{2}} +\alpha^{2} (\beta - n)^{2}
\label{r50}.
\end{equation}
The Hamiltonian in this case is non-Hermitian or pseudo-Hermitian, this has interesting consequences in random matrix theories \cite{AhJ03} and the decay bound state or decay resonances \cite{Zno041,Zno042, BeB98, FeE99}. The loss of hermiticity of the Hamiltonian (in this case) seems to imply complex energies for localized eigenstates, with Im $E_{n} \neq 0$ and consequently a certain rate of decay of any localized state \cite{FeE99}.

The upper and lower spinor components $F_{n, -1}$ and $G_{n, -1}$ of the radial wavefunctions corresponding to this energy equation are:   
\begin{equation}
F_{n, -1} (r)=  N_{n, -1}(-1)^{\lambda}e^{-2\alpha \lambda r}(1 + e^{-2\alpha \lambda r})^{\beta}~~ _{2}F_{1}(2 \beta + n, 2 \lambda - n, 2 \lambda + 1; -e^{-2\alpha \lambda r})
\label{r51},
\end{equation}
and
\begin{eqnarray}
&\displaystyle{G_{n, -1} (r)= -N_{n, -1}(-1)^{\lambda}\frac{e^{-2\alpha \lambda r}(1 + e^{-2\alpha r})^{\beta}}{M + E_{n, -1} - C}}  \\ \nonumber
& \times [  \left( 2 \alpha \lambda + 2 \alpha \beta e^{-2\alpha r}(1 + e^{-2\alpha r})^{-1}  + \frac{1}{r}\right)
 ~_{2} F_{1}(2 \beta + n, 2 \lambda - n, 2 \lambda + 1; -e^{-2\alpha r})\\ \nonumber
& - \frac{2 \alpha (2 \beta + n)(2 \lambda - n) e^{-2\alpha r}}{2 \lambda + 1}~~ _{2}F_{1}(2 \beta + n + 1, 2 \lambda - n + 1, 2 \lambda + 2; -e^{-2\alpha r}) ]
\label{r52},
\end{eqnarray}
where parameters $\lambda$ and $\beta$ become
\begin{equation}
\lambda = \frac{1}{2}[n + \beta - i \frac{(M + E_{n, -1}  - C_{s})V_{2}}{2 \alpha^{2} (\beta - n)}],
\label{r53}
\end{equation}
and
\begin{equation}
\beta = \frac{1}{2}\left(-1 + \sqrt{1 + \frac{4 V_{1}(M + E_{n, -1} - C_{s}) }{\alpha^{2}}}\right)
\label{r54}.
\end{equation}

For the case of the pseudospin symmetry solutions of this potential, $E_{n, 1}$, $G_{n, 1}$ and $F_{n, 1}$ are respectively obtained as :   
\begin{equation}
M^{2} - E_{n, 1}^{2}  + C_{ps} (M + E_{n, 1}) = - \frac{(M - E_{n, 1}  + C_{ps})^{2}V_{2}^{2}}{4 \alpha^{2} (\beta - n)^{2}} +\alpha^{2} (\beta - n)^{2}
\label{r55},
\end{equation}
\begin{equation}
G_{n, 1} (r)= N_{n, 1}(-1)^{\lambda}e^{-2 \alpha \lambda r}(1 + e^{-2\alpha  r})^{\beta}~~ _{2}F_{1}(2 \beta + n, 2 \lambda - n, 2 \lambda + 1; -e^{-2\alpha r})
\label{r56},
\end{equation}
\begin{eqnarray}
&\displaystyle{F_{n, 1} (r)= N_{n, 1}(-1)^{\lambda}\frac{e^{-2\alpha \lambda r}(1 + e^{-2\alpha r})^{\beta}}{M - E_{n, 1} + C_{ps}}}  \\ \nonumber
& \times [  \left( -2 \alpha \lambda + 2 \alpha \beta e^{-2\alpha r}(1 + e^{-2\alpha r})^{-1}  + \frac{1}{r}\right)
 ~_{2} F_{1}(2 \beta + n, 2 \lambda - n, 2 \lambda + 1; -e^{-2\alpha r})\\ \nonumber
& - \frac{2 \alpha (2 \beta + n)(2 \lambda - n) e^{-2\alpha r}}{2 \lambda + 1}~~ _{2}F_{1}(2 \beta + n + 1, 2 \lambda - n + 1, 2 \lambda + 2; -e^{-2\alpha r}) ]
\label{r57},
\end{eqnarray}
where $\lambda$ and $\beta$ are given as, 
\begin{equation}
\lambda = \frac{1}{2}[n + \beta + i \frac{(M - E_{n, 1} + C_{ps})V_{2}}{2 \alpha^{2} (\beta - n)}]
\label{r58}
\end{equation}
and 
\begin{equation}
\beta = \frac{1}{2}\left[-1 + \sqrt{1 - \frac{4 (M - E_{n, 1} + C_{ps})V_{1} }{\alpha^{2}}}\right]
\label{r59}.
\end{equation}

\subsection{Reflectionless-type potential}

Choosing $V_{2} = 0$ and $V_{1} = \frac{1}{2} \gamma (\gamma + 1)$ in the eq. (\ref{r1}), then, eq. (\ref{r1}) becomes the reflectionless-type potential\cite{GrS98,ZhE05}.
\begin{equation}
V(r)= - \frac{}{}\gamma(\gamma + 1)  \textrm{sech}^{2}{\alpha r}, 
\label{r60}
\end{equation}
where $\gamma$ is an integer, {\it i. e.} $\gamma = 1, 2, 3, \ldots, \ldots$~~~~~~.

For the spin symmetry solutions of the reflectionless-type potential, the energy eigenvalues, $E_{n, -1}$ is, 
\begin{equation}
M^{2} - E_{n, -1}^{2}  - C_{s} (M - E_{n, -1}) = \alpha^{2} (\beta - n)^{2}
\label{r61}.
\end{equation}
The upper and lower spinor components $F_{n, -1}$ and $G_{n, -1}$ of the radial wavefunctions corresponding to this energy level are given, respectively as :   
\begin{equation}
F_{n, -1} (r)=  N_{n, -1}(-1)^{\lambda}e^{-2\alpha \lambda r}(1 + e^{-2\alpha \lambda r})^{\beta}~~ _{2}F_{1}(2 \beta + n, 2 \lambda - n, 2 \lambda + 1; -e^{-2\alpha \lambda r})
\label{r62},
\end{equation}
\begin{eqnarray}
&\displaystyle{G_{n, -1} (r)= -N_{n, -1}(-1)^{\lambda}\frac{e^{-2\alpha \lambda r}(1 + e^{-2\alpha r})^{\beta}}{M + E_{n, -1} - C}}  \\ \nonumber
& \times [  \left( 2 \alpha \lambda + 2 \alpha \beta e^{-2\alpha r}(1 + e^{-2\alpha r})^{-1}  + \frac{1}{r}\right)
 ~_{2} F_{1}(2 \beta + n, 2 \lambda - n, 2 \lambda + 1; -e^{-2\alpha r})\\ \nonumber
& - \frac{2 \alpha (2 \beta + n)(2 \lambda - n) e^{-2\alpha r}}{2 \lambda + 1}~~ _{2}F_{1}(2 \beta + n + 1, 2 \lambda - n + 1, 2 \lambda + 2; -e^{-2\alpha r}) ]
\label{r63},
\end{eqnarray}
the parameters $\lambda$ becomes
\begin{equation}
\lambda = \frac{1}{2}[n + \beta ],
\label{r64}
\end{equation}
where
\begin{equation}
\beta = \frac{1}{2}\left(-1 + \sqrt{1 + \frac{2\gamma(\gamma + 1)(M + E_{n, -1} - C_{s}) }{\alpha^{2}}}\right)
\label{r65}.
\end{equation}

Similarly, the pseudospin symmetry solutions of this potential, $E_{n, 1}$,  $G_{n, 1}$ and $F_{n, 1}$ are respectively obtained as :   
\begin{equation}
M^{2} - E_{n, 1}^{2}  + C_{ps} (M + E_{n, 1}) = \alpha^{2} (\beta - n)^{2}
\label{r66},
\end{equation}
\begin{equation}
G_{n, 1} (r)= N_{n, 1}(-1)^{\lambda}e^{-2 \alpha \lambda r}(1 + e^{-2\alpha  r})^{\beta}~~ _{2}F_{1}(2 \beta + n, 2 \lambda - n, 2 \lambda + 1; -e^{-2\alpha r})
\label{r67},
\end{equation}
\begin{eqnarray}
&\displaystyle{F_{n, 1} (r)= N_{n, 1}(-1)^{\lambda}\frac{e^{-2\alpha \lambda r}(1 + e^{-2\alpha r})^{\beta}}{M - E_{n, 1} + C_{ps}}}  \\ \nonumber
& \times [  \left( -2 \alpha \lambda + 2 \alpha \beta e^{-2\alpha r}(1 + e^{-2\alpha r})^{-1}  + \frac{1}{r}\right)
 ~_{2} F_{1}(2 \beta + n, 2 \lambda - n, 2 \lambda + 1; -e^{-2\alpha r})\\ \nonumber
& - \frac{2 \alpha (2 \beta + n)(2 \lambda - n) e^{-2\alpha r}}{2 \lambda + 1}~~ _{2}F_{1}(2 \beta + n + 1, 2 \lambda - n + 1, 2 \lambda + 2; -e^{-2\alpha r}) ]
\label{r68},
\end{eqnarray}
the parameter $\lambda$ is given as, 
\begin{equation}
\lambda = \frac{1}{2}[n + \beta]
\label{r69}
\end{equation}
where
\begin{equation}
\beta = \frac{1}{2}\left[-1 + \sqrt{1 - \frac{2\gamma ( \gamma + 1) (M - E_{n, 1} + C_{ps}) }{\alpha^{2}}}\right]
\label{r70}.
\end{equation}

\section{Conclusion}
In conclusion, we obtained the energy eigenvalues and the spinor wavefunction of the Dirac equation with the Rosen-Morse potential  for $s$-wave bound states under the spin symmetry and pseudospin symmetry. By adopting the basic concepts of the supersymmetric quantum mechanics approach and function analysis method, the energy eigenvalues and associated two-components spinors are obtained. These solutions have been obtained analytically. With appropriate values of $V_{1}$ and $V_{2}$, the spin symmetry and pseudospin symmetry solutions (the energy equations and the corresponding upper and lower spinor components) of the Eckart-type potential, PT-symmetric version of the Rosen-Morse potentials and the reflectionless-type potential are obtained. We suggest that with this approach, $\kappa$-state solutions of the Dirac equation for the Rosen-Morse potential with pseudospin and spin symmetry can be obtained using improved approximate analytical approach, that is, for case $\ell \neq 0$, problems with centrifugal/pseudo-centrifugal term. It is worth noting that at the final preparation of this work, it was found that using improved approximate analytical approach for case $\ell \neq 0$, the solutions of the Dirac equation for the Rosen-Morse potential with pseudospin symmetry and spin symmetry have been obtained using Nikiforov-Uvarov method \cite{IkS10}.\\

{\bf Aknowledgements}

KJO is very grateful to the ICTP for his postdoctoral position at the ICMPA-UNESCO CHAIR
under the Prj -$15$ ICTP project, this work is a product of the motivations I received
from Prof. M. N. Hounkonnou (The Chair, ICMPA, Universite d'Abomey-Calavi, Benin) and
other ICMPA-UNESCO chair visiting Professors. He acknowledges the receipt of some papers on PT- symmetric quantum mechanics from Prof. M. Znojil in $2004$. Also, we appreciate the efforts of
 Profs.  Ginocchio, J. N.; Berkdermir, C.; Soylu, A.; Jia, C. S. and Emeritus Prof. K. T. Hecht for sending valuable materials to us. We thank the anonymous referees for their helpful comments that improved the quality of the paper.

{\it A preliminary report on this work was contributed to the Sixth International Workshop on Contemporary Problems in Mathematical Physics (COPROMAPH 6); under the auspices of International Chair in Mathematical Physics and 
	Applications (ICMPA), Universite d'Abomey Calavi, Cotonuo, Benin Republic (24th Oct. - 6th Nov. 2009).}

%Soylu, A. , Bayrak, O. and Boztosun, I. (2008): J. Phys. A: Math. Theor. {\bf 41}, 065308.\\

\end{document}